# Spin-polarized DFT calculations for physical properties of novel KVSb half-Heusler compound for spintronic and thermodynamic applicability

Ashwani Kumar[1], Anupam[2], Shyam Lal Gupta[3], Sumit Kumar[4], Vipan Kumar[5] and Diwaker Kumar[6*]

[1]Department of Physics, Abhilashi University Mandi-175045 (INDIA)
[2]Department of Physics, RG Govt. Degree College Jogindernagar-175015 (INDIA)
[3]Department of Physics, Harish Chandra Research Institute, Allahabad- 211019 (INDIA)
[4]Department of Physics, Govt. Degree College Una- 174303 (INDIA)
[5,6]Department of Physics, SCVB Govt. Degree College Palampur- 176061 (INDIA)

*Email-diwaker1physics@gmail.com*

## Abstract

In the reported study we have investigated the robust phase stability, elasto-mechanical, thermophysical and magnetic properties of KVSb half Heusler compound by implementing density functional theory models in Wien2k simulation package. The dynamic phase stability is computed in phase type I, II & III phase configurations by optimising their energy. It is observed that given compound is more stable in spin-polarised state of phase type I. To explore the electronic band structure, we apply the generalised gradient approximation. The electronic band profile of the Heusler alloy display a half-metallic nature. Moreover, the calculated second-order elastic parameters divulge the ductile nature. To understand the thermodynamical and thermoelectric stability of the alloy at various temperature and pressures ranges we have utilised the Quasi-Harmonic Debye model. The computed value of magnetic moment found in good agreement with Slater-Pauling rule. Our findings confirms that the predicted half Heusler alloy can be used in various spintronics and thermoelectric applications.

**1. Introduction:** In the recent years, researchers around the globe have revealed extensive interest in spintronic materials owing to their wide array of applications in numerous devices [1-4]. Spintronics is a growing field which come into view on manipulating the electron spin properties to achieve the high-speed information processing and large storage [5]. Devices based on spintronic technology are becoming indispensable part of modern solid-state devices which are based upon both spin and mobility of charge carriers [6-9]. Among these futuristic devices, the assimilation of spin injection and single electron spin sources emerged as one of the most significantly studied application. The spin polarized charge carriers stand out as a corner stone for modern spintronic devices. Due to the usefulness and prominence the Heusler compounds have stand out as frontrunners in this domain [10-13]. Heusler compounds are the ideal spin-polarized materials [14-17] which has created strong path way to tailor the novel devices. The materials with spin-polarisation are extensively important in rapidly processing of the massive data and storing information [18-21]. Heusler compounds exhibit spin dependent electronic properties which is popularly known as half metallic behaviour where the materials show metallic nature for one spin (i.e. spin or spin down) channel and semiconducting for the other (spin up or down) channel consequently possess 100% spin polarisation [22, 23]. The density of states profile for spin up and down channels configuration at the Fermi level ($E_F$) is used to estimate the degree of spin polarisation [24].

Thermoelectricity (TE) is a technology known for the thermoelectric effect which convert heat into electric current and vice versa without hampering the environment. It is one of robust source of clean energy. Thermoelectric technology may be incorporated to reuse the waste heat evolved from automobiles, refineries, other industrial projects and significantly may contribute in the objective to achieve zero carbon emission. Heusler compounds stand out as potential thermoelectric candidates as they are cost effective, easy to synthesize, environmentally compatible and possess stable crystal structure. These characteristics place them in the front row. Based on structural stoichiometry the Heusler compounds are categorized mainly in three types viz. half Heusler (HH), full Heusler (FH) and quaternary Heusler (QH).

Over the years, numerous researchers have conducted both theoretical and experimental studies to explore the spintronic and thermoelectric properties of Heusler compounds [25-30]. Groot et al. [31] were the first to report experimental research on the half-metallic behavior of NiMnSb half-Heusler compound. Shortly after this $Fe_2TiSb$ and $Fe_2TiAs$ were reported as half metals full Heusler compounds [32]. Luo and colleagues reported theoretical work on the half-metallic properties of the full Heusler compounds $Mn_2FeGa$, $Mn_2FeAl$, $Mn_2FeGe$, and $Mn_2FeSe$ [33]. Among the diverse range of Heusler compounds, the half Heusler alloys stand out as an excellent material [34] differentiated by its structural portfolio.

Transitioning to the HH from FH, they are generally relaxed in generic formula XYZ, where X and Y are metal atoms from either the transition elements or the alkaline earth family whereas, Z are mainly from sp block element. The structural composition of HH exhibit the a $C_{1b}$ crystal structure and comprises of three interpenetrating *fcc* lattices which differ from full Heuslers by having one vacant position within their structure.

With multifaced growth in spintronic applications, various HH alloys have been previously reported as versatile spintronic and thermoelectric material such as XCrSb (X=Fe, Ni) [35], KCrZ (Z=S, Se, Te) [36], NaZrZ (Z=P, Sb, As) [37], LiMnZ (Z=N, Si, P) [38], RhCrZ (Z=Ge, Si) [39], PtXBi (Co, Fe, Mn, Ni) [40], IrCrZ (Z=Ge, Sn, As, Sb) [41] and VCoSb [42]. HH alloys are the materials with 100% spin polarisation also known as half-metallic ferromagnets (HMFs) [43]. The HMFs have versatile applications in spin injection [44], magnetic tunnel junction (MTJ) [45], spin valve devices [46], tunnelling magnetoresistance (TMR) [47], magnetic random-access memory (MRAM) [48] and permanent magnets [49].

Despite the extensive range of documented HAs, there is an ongoing requirement for further half-metallic materials to improve the proficiency of spintronic and thermoelectric devices. Potassium (K) based half Heusler compounds, distinguished for their structural robustness and potential for half-metallic ferromagnetism (HMFs) offers possibilities in TE and spintronic devices fabrication.

The aim of proposed work is to perform a comprehensive analysis of physical properties of KVSb half-Heusler compound by utilizing density functional theory models in conjunction with Wien2k simulation program package. This study emphasis to investigate the multiple facets including structural phase stability, half-metallic ferromagnetism (HMF), mechanical stability, magnetic properties and thermodynamic response of KVSb HH compound. The research aims to determine the most stable magnetic configuration, evaluate the half metallic ferromagnetism, $2^{nd}$ order elastic parameters, magnetic moment calculation, Curie temperature and the electronic properties using the density functionals PBE-GGA approximation. At the same time a detailed investigation on the thermodynamic performance under temperature and pressure variations have been put forward using the BoltzTraP code and quasi-harmonic Debye approximation. Moreover, the main intention is to provide a thorough computational detail to the fundamental properties of proposed HH compound and its versatile applications in sustainable spintronic, electronic, thermoelectrics and other emerging fields.

## 2. Crystal structure and computational detail

In the present work we applied density functional theory (DFT) to explore the dynamical structural stability, half-metallic behaviour, magnetic properties and thermodynamical response of alkaline earth metal based half Heusler KVSb compound. Renowned for its precision and reliability DFT ensures the high accuracy in the obtained results. The full potential linearly augmented plane wave (FP-LAPW) basis set was successfully implemented within the WIEN2k simulation package [50-56] to resolve the Kohn-Sham equations. The electron-electron interaction was excellently handled within the generalizes gradient approximation precomputed by Perdew-Burke-Ernzerhof (PBE) exchange-correlation potential [57]. To obtain the detailed electronic structure, modified Becke Johnson (mBJ) potential is adopted and structural stability, magnetic properties are investigated and analysed successfully. The simulations are initiated by dividing structural unit cell of KVSb in two sections one is muffin-tin sphere and other is interstitial space. The muffin-tin is a prominent approximation method generally utilised to estimate the energy state for an electron within the crystal lattice. The value of muffin-tin radii ($R_{MT}$) for K, V and Sb atoms are 2.23, 2.18 & 2.25 respectively. Outside the sphere wave functions of the electrons are expressed in the form of plane waves with the wave vector cut-off of $R_{MT} \times K_{max} = 8.0$, here $K_{max}$ signifies the reciprocal lattice vector. To differentiate the core and valence electrons a fixed energy of magnitude -7.0 Ry is used and to attain the scf convergence a fixed energy value of $10^{-5}$ Ry is utilised successfully. The Brillouin zone for the system is sampled by executing the Monkhorst pack [58] technique with a (14 × 14 × 14) k-point mesh. To investigate the elastic parameters of KVSb IRelast in WIEN2K package is implemented. The thermodynamic properties are explored with BoltzTrap2 code [59] implemented in classical Boltzmann approximation.

## 3. Results and discussions

### 3.1. Prediction of crystal structure

To predict the properties of any material it is essential to understand its crystal structure first. Half Heusler possess $C_{1b}$ type crystal structure and relaxed in three dissimilar atomic configurations. These different structures are shown in **Fig. 1**. The three crystal structure types are signified by distinct atomic co-ordinates and Wyckoff's position are shown as, type I): K (1/4, 1/4, 1/4 ), V (3/4, 3/4, 3/4), Sb (0, 0, 0); type II): K(0, 0, 0), V (3/4, 3/4, 3/4), Sb (1/4, 1/4, 1/4 ) and type III): K (1/2, 1/2, 1/2), V(3/4, 3/4, 3/4), Sb (0, 0, 0). The detailed Wyckoff's positions for type I, II and III KVSb half Heusler are given in the **Table 1**.

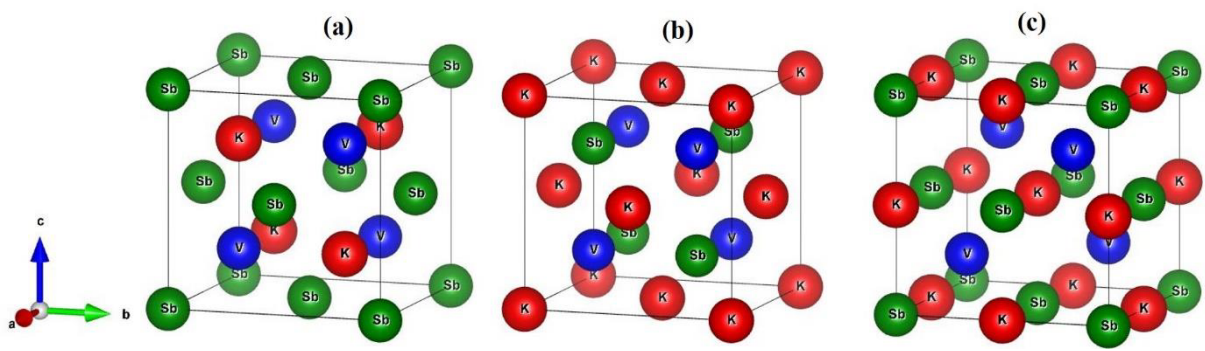

**Fig. 1:** Crystal structures of KVSb half Heusler compound in three different phase arrangements (a) type I (b) type (II) and (c) type (III)

**Table 1:** Wyckoff's atomic sites for KVSb half Heusler alloy

| Phase | K | V | Sb |
|---|---|---|---|
| Type I | (0.25, 0.25, 0.25) | (0.75, 0.75, 0.75) | (0, 0, 0) |
| Type II | (0,0,0) | (0.75,0.75,0.75) | (0.25, 0.25, 0.25) |
| Type III | (0.50, 0.50, 0.50) | (0.75, 0.75, 0.75) | (0, 0, 0) |

It is observed that the dynamical stability of the proposed HH is depends upon the atomic positions in the given crystal structure. So, it is inevitable to identified the most stable phase among the existing phases which have minimum energy. On the basis of calculations performed that type I structure of KVSb HH found to be most stable as compared to type II and III. Henceforth, KVSb with space group *F-43m* (#216) is more dynamical stable in the spin polarised (SP) phase than the non-magnetic (NM). As specified in **Table 2**, we find the equilibrium lattice parameters and minimum energy for cubic KVSb with the stable type-I crystal structure by applying the Murnaghan equation of states [60] are expressed in equation (1) for the volume optimization in spin polarized (SP) phase.

$$E_T(V) = E_0 + \frac{B_0 V}{B'_0(B'_0-1)}\left\{\left(\frac{V_0}{V}\right)^{B'_0} - 1\right) + B'_0\left(1 - \frac{V_0}{V}\right)\right\} \quad (1)$$

where $E_0$ signifies the minimum total energy, $V_0$ is volume of the unit cell at zero pressure, $B_0$ and $B'$ are the bulk modulus and derivative of bulk modulus respectively. As delineate in **Table 2** the computed lattice constant ($a_0$) for KVSb HH at symmetry as obtained with GGA is 6.21Å. The other parameters $V_0$, $B_0$ and $B_o'$ are reported first time for KVSb HH alloy.

**Table 2:** Computed structural dynamic parameters.

| Alloy | Type | $a_0$(Å) | $B_0$(GPa) | Volume (a.u.$^3$) | Energy (Ry) | |
|---|---|---|---|---|---|---|
| | | SP | SP | SP | SP | NM |
| KVSb | I | 6.99 | 34.96 | 577.39 | -16070.01 | -16069.95 |
| KVSb | II | 6.98 | 35.23 | 578.40 | -16071.10 | -16069.21 |
| KVSb | III | 6.96 | 35.54 | 578.87 | -16072.11 | -16069.22 |

A total energy minimization approach is implemented to find the most stable magnetic configuration. For this we have calculated the volume and energy curve and found that spin-polarized (SP) state in phase I is more favourable than the non-magnetic (NM) for the proposed KVSb HH compounds. The energy-volume data is fit into the Murnaghan equation of state as shown in eqn. 1. The lattice constants for all three states are computed in the proposed work,

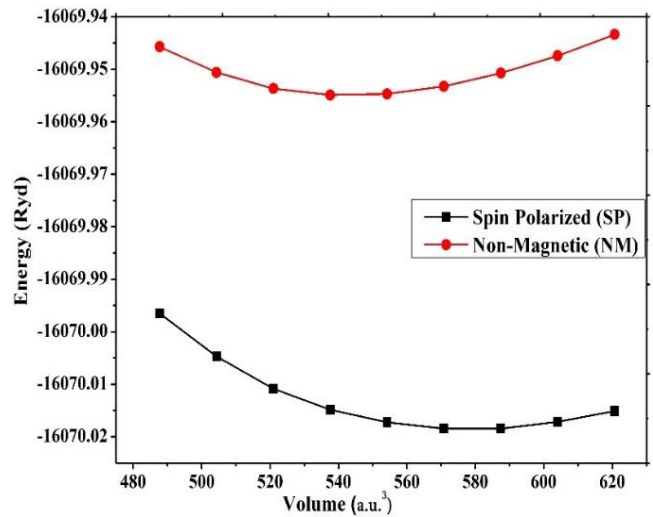

**Fig. 2:** Optimised energy versus volume curve for spin polarised crystal structure of cubic KVSb HH in stable type 1 form.

## 3.2 Mechanical and elastic properties

The elastic and mechanical properties of a material can be determined by calculating the different elastic parameters and to establish relation between the stress and strain. The elastic parameters determine the deformation experienced by the material under the external force and then attains its original configuration. To examine the mechanical stability of compound the elastic constants are very significant to determine at the optimised state. Various practical applications of solids are related to their mechanical attributes and are determined by finding the elastic constants [61]. The KVSb HH is stable in phase I cubic symmetry, so we calculated the values of three independent elastic parameters $C_{11}$, $C_{12}$ and $C_{44}$ respectively. Where $C_{11}$ measures the longitudinal expansion and $C_{12}$ measures the transverse deformation. To defines the mechanical stability the independent elastic constants must fulfil the Born's stability criteria which is given for cubic symmetry as-

$$C_{11} - C_{12} > 0 \text{ and}$$
$$C_{11} + 2C_{12} > 0$$
$$\text{Where, } C_{11} > 0 \text{ ; } C_{44} > 0 \tag{2}$$

**Table 3** shows that KVSb HH compound meet the stability criteria and indicating its mechanical stability firmly. In addition to this the other listed parameters titled shear modulus (G) and bulk modulus (B) which determines the compressibility and stiffness. The Resistance to the plastic deformation, which measures the opposing force to fracture, can be determined by the parameters G and B. These values may be obtained by using the eqns. given as-

$$B = \frac{C_{11} + 2C_{12}}{3} \tag{3}$$

$$\text{and} \quad G = \frac{G_R + G_V}{2} \tag{4}$$

The R and V in the subscript of Eqn. 4 signifies the Reuss and Voigt bounds and the values of $G_R$ and $B+G_v$ can be determines by using the following relations-

$$\textit{i.e.} \quad G_R = \frac{5C_{11}(C_{11} - C_{12})}{5C_{11} + 3(C_{11} - C_{12})} \tag{5}$$

$$\text{and} \quad G_V = \frac{C_{11} - C_{12} + 3C_{44}}{5} \tag{6}$$

The computed values of the above-mentioned elastic parameters are shown in **Table 3-**

**Table 3:** Computed values of elastic parameters ($C_{ij}$), bulk and Young's modulus (B&E), Cauchy pressure (CP), Poisson's ratio (v), Voigt shear modulus ($G_v$), Reuss shear modulus ($G_R$), Shear modulus (G), Pugh's ratio (B/G), Kleinman's parameter($\xi$), transverse, longitudinal and average wave velocity ($v_A$, $v_l$ & $v_t$), Debye temperature ($\theta_D$), Chen-Vickers hardness ($H^{CV}$), Tian-Vickers hardness ($H^{TV}$), Lame's 1st & 2nd parameter ($\lambda$ & $\mu$) under external pressure of 0, 5 & 10 GPa for cubic KVSb HH alloy.

| Stress | KVSb HH alloy | | |
|---|---|---|---|
| Parameters | 0 GPa | 5 GPa | 10 GPa |
| $C_{11}$ | 503.318 | -169.315 | 3805.812 |
| $C_{12}$ | -247.721 | -613.131 | -1855.655 |
| $C_{11}-C_{12}$ | 751.040 | 443.816 | 5661.467 |
| $C_{11}+2C_{12}$ | 7.875 | -1395.579 | 94.502 |
| $C_{44}$ | 371.790 | 217.934 | 2836.460 |
| Voigt bulk modulus($B_v$) | 2.625 | -465.192 | 31.501 |
| Reuss bulk modulus($B_R$) | 2.625 | -465.192 | 31.501 |
| Hill bulk modulus($B_H$) | 2.625 | -465.192 | 31.501 |
| Voigt shear modulus($G_S$) | 373.282 | 219.524 | 2834.169 |
| Reuss shear modulus($G_v$) | 373.273 | 219.506 | 2834.167 |
| Hill shear modulus($G_H$) | 372.277 | 219.515 | 2834.168 |
| Voigt Young's modulus($E_Y$) | 23.140 | 781.500 | 274.358 |
| Reuss Young's modulus($E_R$) | 23.140 | 781.428 | 274.358 |
| Hill Young's modulus($E_H$) | 23.140 | 781.464 | 274.358 |
| Voigt Poisson's coefficient($v_v$) | -0.969 | 0.780 | -0.952 |
| Reuss Poisson's coefficient($v_R$) | -0.969 | 0.780 | -0.952 |
| Hill's Poisson's coefficient($v_H$) | -0.969 | 0.780 | -0.952 |
| Kleinman's parameter ($\xi$) | -0.368 | -123.593 | -0.364 |
| Transverse elastic wave velocity ($v_t$) | 9529.727 m/s | 6903.724 m/s | 23877.243 m/s |
| Longitudinal elastic wave velocity ($v_l$) | 11032.966 m/sec | ---- | 27685.743 m/s |
| Average wave velocity ($v_A$) | 9939.043 m/sec | ---- | 24912.016 m/s |
| Debye temperature($\theta_D$) | 968.487 K | ---- | 2586.349 K |
| Pugh's ratio(K) | 0.007 | -2.119 | 0.011 |
| Chen-Vickers hardness($H^{CV}$) | 21107.550 | 16.461 | 40454.836 |
| Tian-Vickers hardness($H^{TV}$) | 17083.364 | ---- | 42650.433 |
| Lame's 1st parameter ($\lambda$) | -246.226 | -611.536 | -1857.945 |
| Lame's 2nd parameter ($\mu$) | 373.277 | 219.515 | 2834.168 |

The value of bulk modulus (B) obtained from eqn.3 is same given in the table 3. The value of Young's modulus which is the ratio of stress to strain is calculated by using the relation-

$$E = \frac{9GB}{3G+B} \quad (7)$$

To check the brittle and ductile nature of the material Cauchy pressure (CP) relation is most prominent and is given as-

$$CP = C_{12} - C_{44} \quad (8)$$

If the calculated value of Cauchy pressure (CP) is positive in accordance with the eqn, (8) then the material is ductile in nature whereas the negative CP value signifies the brittle character our result confirms that the predicted material is of brittle nature. Pugh ratio (B/G) is another parameter which also define the ductility or brittleness of materials. A material with B/G ratio less than 1.75 are brittle. This obtained B/G ratio confirms that the material is brittle. The computed value of Pugh's ratio for KVSb HH is illustrated in **Table 3**. To check whether the given compound is compressible or incompressible the Poisson's ratio is imperative to calculate. This ratio also imperative to specify the central forces and bonding mechanism within the compound and hence calculated as-

$$\nu = \frac{3B-2G}{2(3B+G)} \quad (9)$$

The threshold value for KVSb alloy clearly describes that it is incompressible in nature. From the elastic of properties exploration, it can be figure out that KVSb HH alloy is ductile and incompressible, specifying their mechanical stability. These parameters are crucial for the practical use of the given alloy. Half-Heusler (HH) alloys fascinating a significant interest for the potential spintronics applications, hence understanding their mechanical stability is very beneficial throughout the device fabrication process.

### 3.3. Electronic properties:

To know the electronic behaviour the band structure and density of states (DOS) are prominent to be studied. In the band structure plot the band gap ($E_g$) determines the thermoelectric response. We have accurately analysed the band gap of KVSb HH alloy by using GGA-PBE scheme. **Fig. 3 (a)** illustrate the electronic structure for spin up configuration which reveals the metallic nature of reported alloy. It means that majority of spin bands are metallic for spin up system. The metallic behaviour for spin up configuration is due to the overlapping of valence band and conduction bands at the Fermi level ($E_F$). Contrary to this, there is band gap in the spin down configuration resulting the semiconducting behaviour for this spin channel is shown in **Fig 3 (b)**. The valence band and conduction band are separated by band gap $E_g$ at the Fermi level.

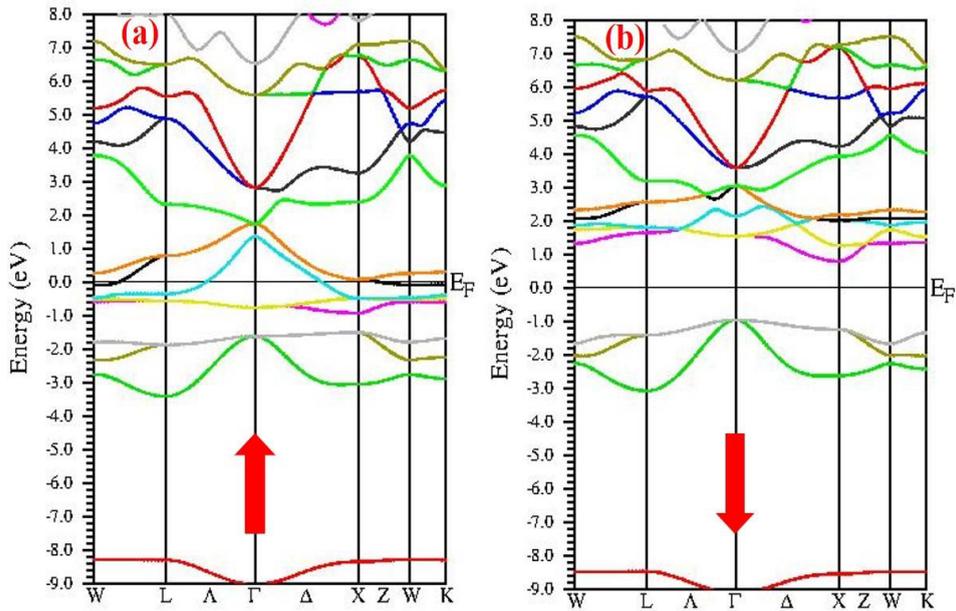

**Fig. 3:** Band structures of KVSb HH compound for a) spin up and b) spin down channels

The deepest point of conduction band is lying along the X symmetry point whereas the highest point of valence band lying at the gamma (Γ) symmetry point from the band structure we may conclude that KVSb exhibit an indirect band gap ($E_g$) of 1.75eV.

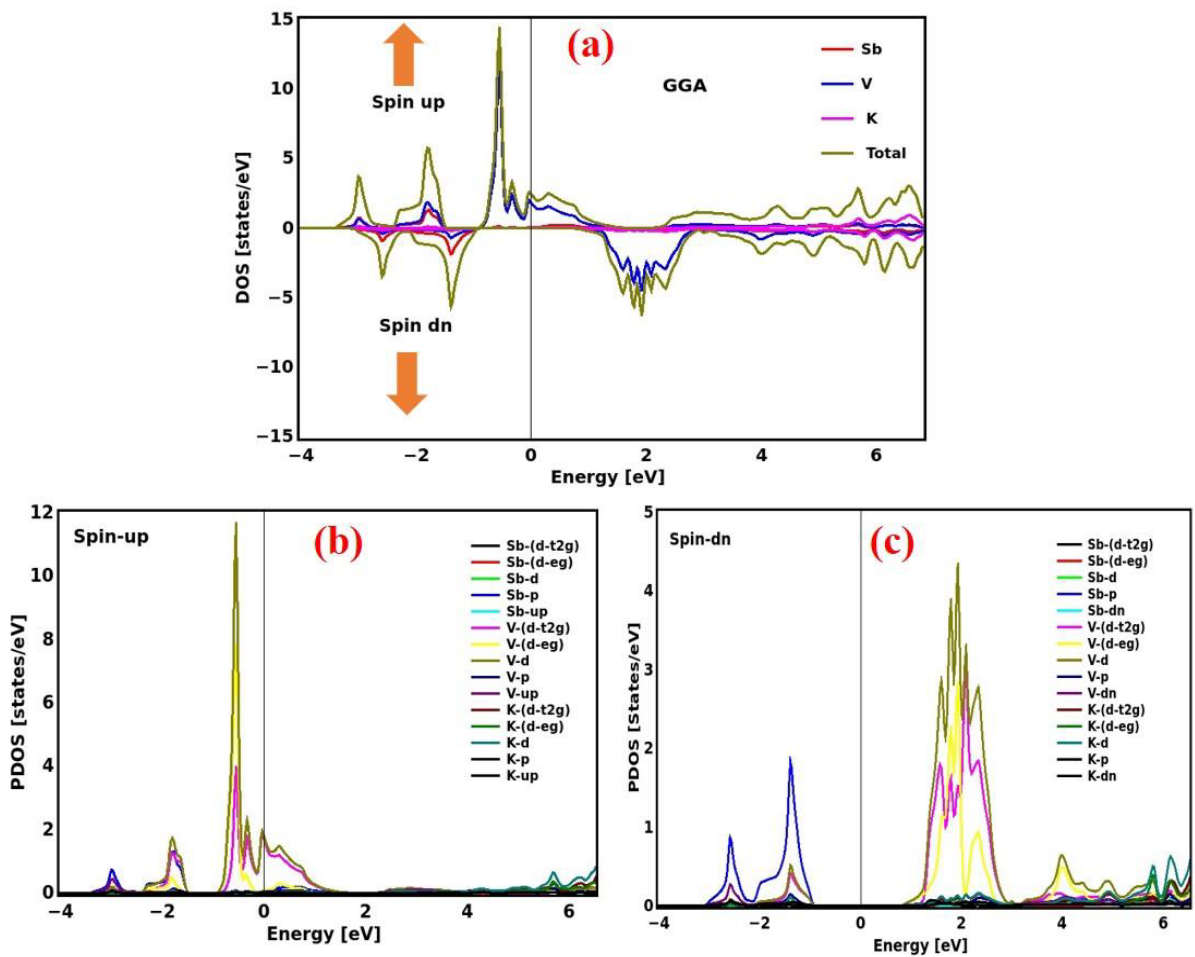

**Fig. 4:** Total and projected density of states (TDOS and pDOS) for a) spin up and (b) spin down channels.

To get the deep insight into electronic attributes the density of state (DOS) plots *i.e,* total (TDOS) and projected density of states (PDOS) for both spiral (spin up and spin down) directions are given in **Fig. 4 (a), (b) & (c)**. The specific effects of K, V, and Sb atoms on the band gap ($E_g$) of given compound is estimated by examining the pDOS. The electronic band structure and DOS plots showed similar trends. Although the p-block Sb atoms had a minor impact on the band gap ($E_g$) formation. The exchange splitting appears between the K and V metal atoms which is more or less responsible for half metallic behaviour.

We observed from pDOS that the majority of DOS near the Fermi energy is mostly due to the *d*-states of V and the p-states of K atoms. Contrary to this, the contributions from the K atoms are from s and p states, whereas Sb atoms s and d states overall contribution is negligible to determine the electronic behaviour. For the both spin channels *p*-states of Sb atoms moderately contributing to the valence band, while the d-states of V atoms make the major role to the majority spin (spin-up channel) and the valence band. In the minority spin state, the d-states of V atoms significantly put out in the conduction band. From the total DOS curves, we use the spin-projected DOS ↑ and DOS ↓ in the spin up or spin down states near the Fermi level ($E_F$) to calculate spin polarization (P), we used the following equation-

$$P = \frac{DOS_\uparrow(E_F) - DOS_\downarrow(E_F)}{DOS_\uparrow(E_F) + DOS_\downarrow(E_F)} \times 100 \qquad (10)$$

$DOS_\uparrow(E_F)$ and $DOS_\downarrow(E_F)$ in Eqn.10 signifies the available states for the majority and minority spin channels at the Fermi level ($E_F$).

### 3.4 Thermodynamic and thermoelectric predictions

Thermodynamic properties are rigorously studied to predict the crucial behaviour of the material for various thermoelectric applications. For this we have implemented the quasi-harmonic Debye model [62] under high temperature and pressure. The impact on specific heat at constant volume ($C_V$), unit cell volume (V), thermal expansion coefficient (α), Debye temperature ($\theta_D$) and entropy change (S) has been investigated across a pressure range from 0-8 GPa and temperature range from 0-500K. The effect of temperature and pressure on unit cell volume is shown in **Fig. 5(a)**. It is manifested that the volume increases with the temperature at a given pressure value. Conversely, with increasing the magnitude of pressure the volume rapidly decreases. Thus, we can conclude that the pressure has a more robust effect on the volume of KVSb material than temperature. The thermal expansion coefficient (α) is another important parameter to check the thermal state of the material. It provides significant information about bonding strength and anharmonicity within the crystal. The change in α with temperature and pressure is illustrated in the **Fig. 5(b)**. It is observed that the value of α increases significantly with the temperature and decreases gradually with increase in pressure.

At absolute zero temperature, the value of α is zero, and it increases swiftly with the temperature and indicating that KVSb follows the Debye's $T^3$ law at lower temperatures.

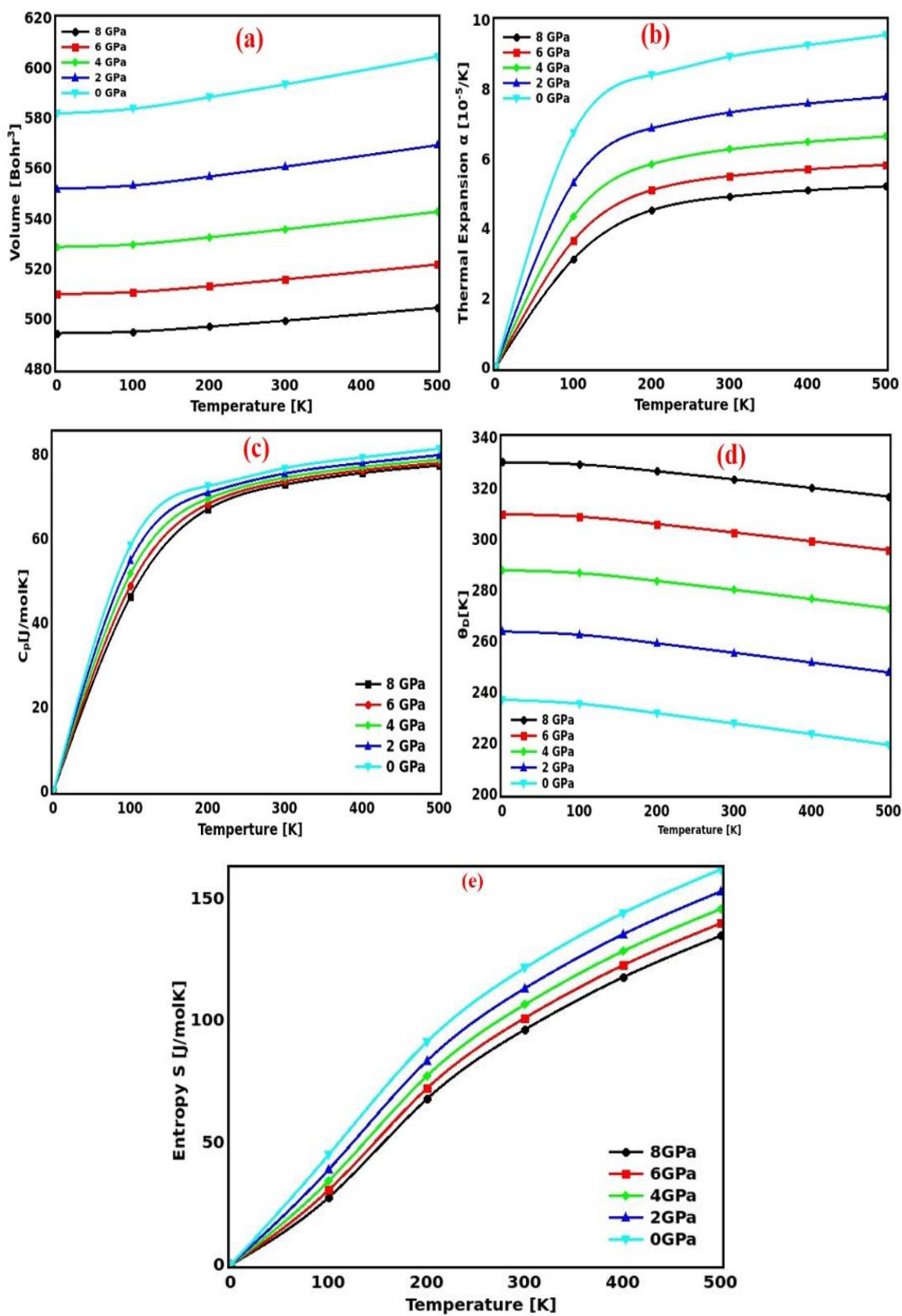

**Fig. 5:** Variation of (a) unit cell volume (b) thermal expansion co-efficient (c) specific heat (d) Debye temperature and (e) entropy with pressure and temperature for KVSb half Heusler compound

The specific heat ($C_V$) as a function of temperature and pressure is computed for KVSb alloy and described in **Fig. 5 (c)**. At low temperature values, the variations of $C_V$ are very close for different pressures resulting the robust dependency on temperature because of the anharmonic effects. At higher temperatures, the specific heat attains a constant value and bow to the Dulong and Petit's rule [63], which is similar for all the materials at higher temperature as a result of anharmonic effect. The behaviour of entropy (S) and Debye temperature ($\theta_D$) as a function of temperature and pressure are also reported. The **Fig. 5 (d)** show that the Debye temperature ($\theta_D$) decreases with increase in temperature confirming the typical behavior generally observed in intermetallic compounds. Though, it displays a rising trend as pressure increases. At constant pressure, the degree at which the Debye temperature declines with growing temperature is comparatively small. The **Fig. 5 (e)** illustrated the variation of entropy (S) with temperature from 0-500K.

To present the electronic transport properties of KVSb HH alloy we used the BoltzTrap code. Seebeck and Peltier effects are the fundamentals of thermoelectric (TE) technology. The efficiency of TE devices is determined by the physical parameter figure of merit (zT) which is expressed as $zT = \frac{\sigma S^2}{\kappa}$, σ and S denotes electronical conductivity, Seebeck coefficient and play significant role to find the efficiency of thermoelectric materials. The thermoelectric materials (TEM) are extensively used in power generators, refrigerators and coolers. The efficiency of TE material is determined by sum of the contributions from electron thermal conductivity ($\kappa_e$) and lattice thermal conductivity ($\kappa_l$) i.e. ($\kappa = \kappa_e + \kappa_l$). A good TE material must have a low thermal conductivity (κ) while maintaining high values for both Seebeck coefficient (S) and electronical conductivity (σ). However, it is challenging to maintain high values for σ and S to obtain appreciable figure of merit (zT). It is due to the fact that electrical conductivity (σ) is inversely and directly related to the effective mass and carrier concentration, whereas the Seebeck coefficient (S) is directly and inversely related to these variables [64, 65]. For many years, these demanding requirements have steer efforts to attain a higher zT. This posing a big challenge to the material researchers and scientist's community in designing the effective TE devices for practical applications. In the reported work, we used the constant relaxation time (τ) approximation to predict TE parameters. This includes parameters such as electrical conductivity (σ/τ), electronic part of thermal conductivity ($\kappa_e$/τ), power factor (PF), figure of merit (zT) and Seebeck coefficient (S). Usually, the electrical conductivity of semiconducting materials increases with rise in temperature because electrons require external energy to overcome the potential barrier, which permit them to travel freely as the temperature rises due to presence of energy gap. For predicted KVSb alloy the value of $\frac{\sigma}{\tau}$ factor decreases with significant increase in temperature as illustrated in **Fig. 6 (a)**. For this factor the computed lowest and highest values are $1.33 \times 10^{20} \Omega m s^{-1}$ and $1.44 \times 10^{20} \Omega m s^{-1}$ respectively. Both lattice ($\kappa_l$) electrical ($\kappa_e$) parts affect the overall thermal conductivity ($\kappa$).

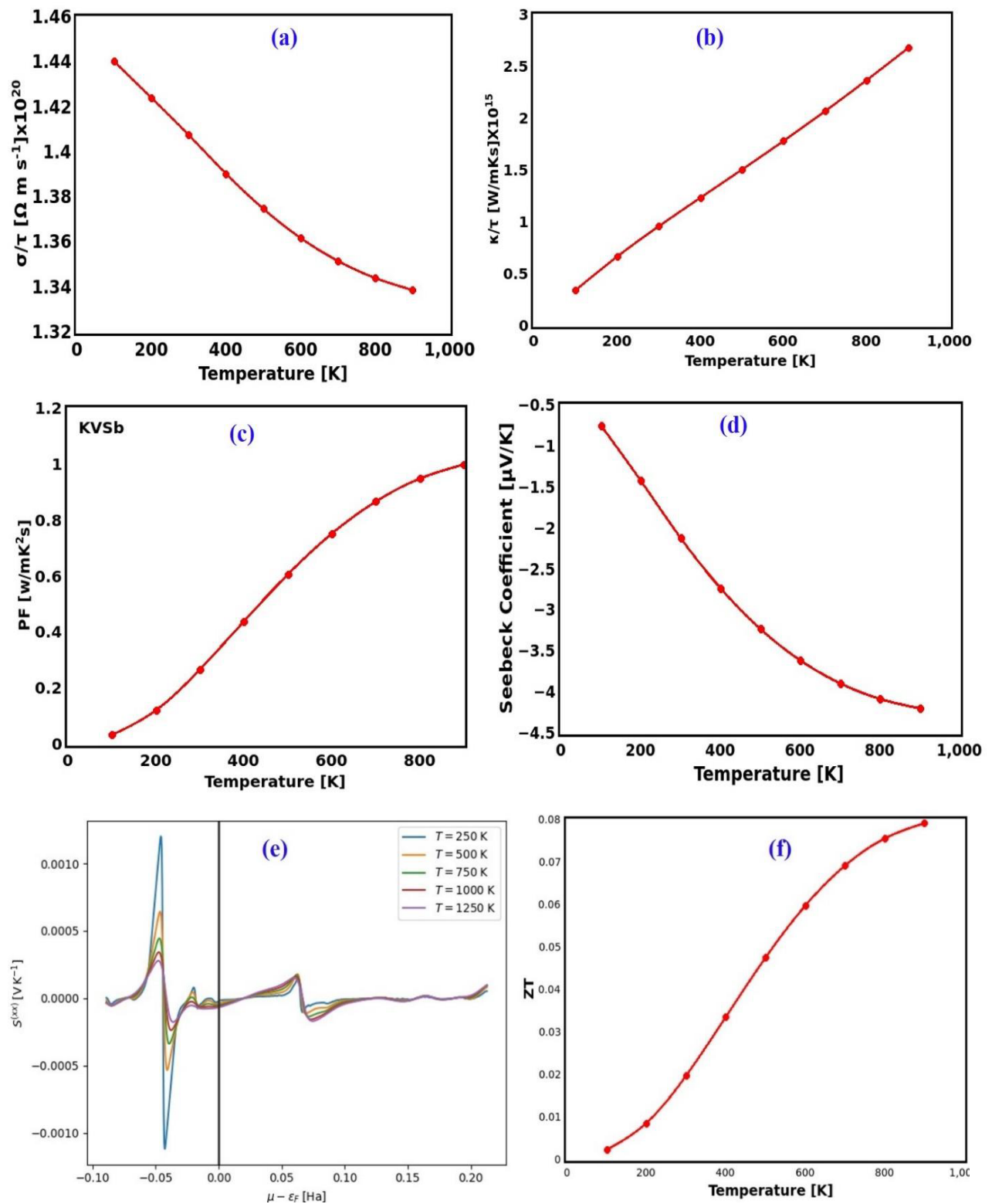

**Fig. 6:** Variation of robust thermoelectric performance for KVSb alloy with temperature and chemical potential

We have predicted the electronic effect of the thermal conductivity only for KVSb system. The approximated value of electronic part of the thermal conductivity exhibits a growing trend with temperature as depicted in **Fig. 6 (b)** for KVSb system. The value of $\kappa_e/\tau$ rises from $0.3 \times 10^{20}$ W/mK at 110 K to $2.7 \times 10^{20}$ W/mK at 975K. The power factor (PF) is an important parameter which is described as $PF = S^2\sigma$. It signifies the appropriateness of a material for TE applications. A higher value of PF specifies a more promising TE material which means it can generate more electrical power against the given temperature gradient. The value of PF increases from 0.1W/mK$^2$s at 120K to 1.0W/mK$^2$s at 900K as shown in **Fig. 6 (c)**. Another important TE parameter is Seebeck coefficient (S). It describes the capacity of the TE material to generate a thermo-emf with the given temperature. The dependency of S with the temperature is given in **Fig. 6 (d)**. Here we can see that value of S for KVSb is -0.1µV/K at 110 K, -2.80 µV/K at 375K and -4.5 µV/K at 945K. All negative values of S explains that the proposed alloy is of n-type semiconductor. In **Fig. 6 (e)** we have also plotted the variation of S with the chemical potential $(\mu - \varepsilon_F)$ in the range of -0.10 eV to 0.20 eV to check the thermoelectric performance. Here we computed the S value at different temperatures i.e., 250K, 500K and 750K respectively. S is notably improved in the locale of µ=0, which signifies that a reasonably large value of S. The maximum value of Seebeck coefficient (S) is $\sim \pm 0.0010$ for both positive and for negative chemical potential. Another parameter we are reporting in **Fig. 6 (f)** is figure of merit (zT). It is a dimensionless quantity which is directly linked to the TE response of materials. The curve exhibits the variation of zT with temperature. The value of zT significantly increase with temperature. At 140 K, the value of zT is 0.01 and 0.08 at 900K. The high value of zT at ambient temperature is particularly noteworthy and significant.

**3.5. Magnetic properties:** The half-metallic nature of a magnetic material can be predicted by using the Slater-Pauling empirical rule [66] and Kubler [67] aligning with the approach forwarded by de Groot et al. [31]. The spin magnetic moment is equal to the difference among occupied energy bands in both spin up and down configurations. The computed magnetic moment of KVSb HH alloy is 3µ$_B$. In **Table 4** the obtained total and partial magnetic moment values per formula unit are given in detail.

**Table 4:** Calculated individual and total magnetic moment for KVSb HH alloy

| Alloy | M$_K$ (µ$_B$) | M$_V$ (µ$_B$) | M$_{Sb}$ (µ$_B$) | M$_{total}$ (µ$_B$) | P (%) | T$_c$ (K) |
|---|---|---|---|---|---|---|
| KVSb | 0.3736 | 2.8117 | -0.2100 | 3.00041 | 100 | 567 |

Our findings indicate that Z atoms (i.e., Sb) do not contribute to the overall magnetic moment. The V atoms are the primary contributors, whereas the contribution from K atoms is minor. The obtained results are in good agreement with SP rule which is expressed as M$_{total}$=Z$_{total}$-8[68], where M$_{total}$ is the total magnetic moment and Z$_{total}$ represent the total number of electrons in the predicted material which is obtained by summing the valence electrons in the three constituent atoms. Curie temperature ($T_c$) is another important parameter of the magnetic materials. Tc is the temperature beyond which the material undergoes phase transition i.e. from magnetic to non-magnetic. At this temperature the intrinsic MM critically changes its directions. The $T_c$ can be determined by applying MM values in a linear relationship [69]-

$$T_c = 23 + 181 M_{total} \qquad (11)$$

Where $T_c$ is the Curie temperature and $M_{total}$ is the total MM. The calculated value of Curie temperature is 567K making the alloy an ideal candidate for high temperature applications.

**4. Conclusion:** In the reported study we have utilised the GGA-PBE approximation to explore the structural dynamics, mechanical, electronic, thermodynamic and thermoelectric performance of the KVSb HH alloy. It was observed that KVSb alloy is stable in type I of its crystal phase and spin polarised phase. The obtained value of lattice constant is 6.99Å, exhibit half-metallic property i.e. metallic for spin up configuration and semiconducting for spin down configuration with an indirect band gap of 1.75 eV. The band structure and density of states profiles supports this and determine that d orbitals of 'V' atoms played an imperative role in overall behaviour. The obtained value of total magnetic moment is of the magnitude 3.00μ$_B$ and satisfies the Slater-Pauling rule. The detailed thermodynamic study is performed by applying Quasi-Harmonic Debye model. The parameters like thermal conductivity, Seebeck coefficient, figure of merit, power factor under temperature limit is successfully investigated and the results provides a fair support to KVSb system. The Elastic calculations advocate that KVSb is brittle, anisotropic incompressible, mechanically and dynamically stable. The reported work provides a strong understanding for the future's experimentalists to make efficient thermoelectric devices based upon KVSb system.